\documentclass[]{emulateapj}
\def\figdir{./}
\newcommand{\ltsima}{\mbox{$\; \buildrel < \over \sim \;$}}
\def\simlt{\lower.5ex\hbox{\ltsima}}
\def\gtsima{$\; \buildrel > \over \sim \;$}
\def\simgt{\lower.5ex\hbox{\gtsima}}

\def\eq{equation}

\def\figname#1{\figdir/#1}

\def\eq{equation}
\def\fig{Fig.}

\def\ie{{\it i.e.}}
\def\eg{{\it e.g.}}
\def\ltsima{$\; \buildrel < \over \sim \;$}
\def\simlt{\lower.5ex\hbox{\ltsima}}
\def\gtsima{$\; \buildrel > \over \sim \;$}
\def\simgt{\lower.5ex\hbox{\gtsima}}
\def\eq{equation}
\def\eqs{equations}

\def\hide#1{}

\def\sig8{$\sigma_8$\xspace}

\def\bi{\begin{itemize}}
\def\ei{\end{itemize}}

\begin{document}

\pagestyle{myheadings} \markright{DRAFT: \today\hfill} 

\title{Bondi accretion in the early universe}

\author{Massimo Ricotti} 
\affil{Department of Astronomy, University of
Maryland, College Park, MD 20742, USA}
\email{ricotti@astro.umd.edu}

\begin{abstract}
This paper presents a study of quasi-steady spherical accretion in the
early Universe, before the formation of the first stars and
galaxies. The main motivation is to derive the basic formulas that
will be used in a companion paper to calculate the accretion
luminosity of primordial black holes and their effect on the cosmic
ionization history.

The following cosmological effects are investigated: the coupling of
the gas to the CMB photon fluid (\ie, Compton drag), Hubble expansion,
and the growth of the dark matter halo seeded by the gravitational
potential of the central point mass.  The gas equations of motion are
solved assuming either a polytropic or an isothermal equation of
state. We consider the cases in which the accreting object is a point
mass or a spherical dark matter halo with power-law density profile,
as predicted by the theory of ``secondary infall''.  Analytical
solutions for the sonic radius and fitting formulas for the accretion
rate are provided.

Different accretion regimes exist depending on the mass of the
accreting object. If the black hole mass is smaller than $\sim
50-100$~ M$_\odot$, gas accretion is unaffected by Compton drag. A
point mass and an extended dark halo of equal mass accrete at the same
rate if $M \simgt 5000$~M$_\odot$, while smaller mass dark halos
accrete less efficiently than the equivalent point mass. For masses $M
\simgt 3 \times 10^4$~M$_\odot$, the viscous term due to the Hubble
expansion becomes important, and the assumption of quasi-steady flow
fails. Hence, the steady Bondi solutions transition to the time-dependent
self-similar solutions for ``cold cosmological infall''.
\end{abstract}

\section{Introduction}

The gravitational collapse of relativistic matter during the early
evolution of the universe is thought to produce primordial black holes
(PBH). Moderately non-linear inhomogeneities that become
gravitationally unstable after entering the particle horizon have a
density and radius that is very close to the black hole regime. It is
easy to show that the Jeans length and the Schwarzschild radius of a
black hole with mean density close to the cosmic value both approach
the particle horizon radius as the effective sound speed of the gas
approaches the speed of light. Hence, small density perturbations with
$\delta \rho /\rho \sim 0.1-1$ are unstable and collapse into PBHs
with a mass that roughly equals the mass within the particle horizon
at the redshift of their formation \citep{Hawking:71, Carr:74,
Musco:05, Harada:05}.

Many physical processes may produce the necessary level of
inhomogeneity for significant PBH formation. In most models PBHs are
produced during phase transitions of the equation of state and,
depending on the time of their formation, may have masses that range
between the Planck mass to $10^5$~M$_\odot$.  Although the probability
of PBH formation is finite, it may be so small to be of little or any
cosmological interest.  PBHs with mass $<10^{15}$ g are thought to
evaporate emitting Hawking radiation on time scales shorter than the
age of the Universe. The radiation or particles they emit would make
them detectable. Thus observational constraints on the abundance of
evaporating PBHs are rather stringent. The upper limit on the density
parameter of PBHs with masses $1 g < M_{PBH} < 10^{15}$ g is $\beta(M)
\sim 10^{-20}-10^{-22}$ at the redshift of formation
\citep[\eg][]{Carr:03}. On the other hand, the existence of
non-evaporating PBHs with mass $>10^{15}$ g is very poorly
constrained. It is conceivable that non-evaporating PBHs constitute
all or a substantial fraction of the dark matter in the Universe. The
existence of PBHs with $M \sim 0.1-1$~M$_\odot$ is of particular
interest because the MACHO collaboration has detected compact objects
in this mass range in the Galactic halo, constituting about 20\% of
the dark matter \citep{Alcock:00}, although this claim is weakened by
recent results \citep[\eg,][]{Hamadache:06}.  PBHs of about
1~M$_\odot$ are also of special interest theoretically because they
may have formed copiously during the quark-hadron phase transition due
to a temporary softening of the cosmic equation of state
\citep{Jedamzik:97}.

The main motivation for the calculations presented this work is to
provide the basic formulas for the accretion rate onto PBHs that,
typically, are thought to have masses smaller than $10^5$~M$_\odot$. We
will show that in this mass range the cosmological spherical accretion
solutions are described by the steady equations (\ie, Bondi solutions)
while for more massive black holes the accretion equations are
time-dependent and the solutions self-similar (\ie, secondary infall
solutions).

These formulas will be the foundation for further calculations
presented in a companion paper \citep[][hereafter,
Paper~II]{RicottiOM:06} aimed to understand whether the interaction
between an, as yet undetected, population of non-evaporating PBHs and
the rest of the matter in the Universe may produce observable
signatures. Thus, in this paper we will only focus on the mathematical
aspect of the spherical accretion problem (\ie, for the mass range we
are interested in, the solution of the steady ``Bondi'' problem)
without a discussion on the physics of the early universe, feedback
effects and other complications. Nevertheless, in the remaining of
this introduction we provide some additional background discussion to
put the mathematical problem into context.

PBHs can accrete gas and dark matter from a nearly uniform medium
\citep{Carr:81, MillerO:01, Mack:06} well before the collapse of the
first nonlinear structures (stars and galaxies). Since the density
inhomogeneities of the gas and dark matter decrease with increasing
redshift, we expect the angular momentum of the accreted material (gas
and dark matter) to be small and hence, the accretion flow to become
increasingly spherical.  At redshifts greater than $z \sim 100$, CMB
photons are strongly coupled to the gas because of Compton scattering
with the residual electrons left over after the epoch of recombination
($x_e \simeq 10^{-4}$).  The gas behaves as a viscous fluid and the
accretion onto PBHs can be substantially reduced. The photon viscosity
may also be important in removing angular momentum from the gas
\citep[\eg,][]{Loeb:93}.  We will show that the Hubble expansion also
acts as a viscous term that tends to reduce the accretion rate,
independent of the ionization fraction of the gas.  Finally, if
massive PBHs do not constitute the bulk of dark matter, their
gravitational potential seeds the formation of a dark halo that is
expected to grow over time becoming $\sim 100$ times more massive than
the PBH in its center \citep{Mack:06}.

The gas accretion onto PBHs produces ionizing radiation and an X-ray
background that may be sufficiently large to heat the cosmic gas and
keep it partially ionized after recombination. Thus, PBHs may imprint
signatures on CMBR which are incompatible with observations by WMAP3
\citep{Spergel:06} and such observations can be used to constrain PBH
masses and abundances.  Several ingredients need to be considered when
modeling ionization by accreting PBHs, including their proper motion
and feedback effects. The X-rays emitted by accreting PBHs, for
example, increase the temperature and fractional ionization of the gas
and are able to reduce the accretion rate. Thus the global accretion
rate onto PBHs and the ionization state of the cosmic gas are coupled
and need to be calculated self-consistently. Such calculations will be
presented in a companion paper (Paper~II).

The present study focuses on the derivation of the basic analytical
relationships for the gas accretion rate as a function of redshift,
temperature, density and ionization fraction of the gas.  We find
solutions for the spherical accretion problem including the following
cosmological effects: i) the effect of the viscous terms due to
Compton drag and Hubble expansion; ii) we consider both the case of an
isolated PBH and the case in which the PBH is ``clothed'' by a dark
matter halo with power-law density profile. We discuss the
relationships between the steady Bondi type solutions and the
time-dependent-self-similar solutions of secondary infall.  Analytical
expressions for the accretion rate as a function of the ``viscosity
parameter'' are provided for either a ``naked'' or ``clothed'' PBH. We
also consider the different cases of isothermal equation of state or
polytropic equation of state with index $\gamma$ for the gas.

\cite{Umemura:94} have previously investigated the effect of radiation
drag on spherical accretion considering three cases for the
gravitational potential: a central point mass, a dark halo with
constant potential and a self-gravitating gas sphere. Their work is
similar to the present work for the case of accretion onto a point
mass. Our work differs from the previous ones in that it addresses the
growth of a dark matter halo around PBHs, it discusses the validity of
the Bondi accretion regime and the transition to time-dependent
solutions. Finally, and most importantly, we derive analytic
expressions for the accretion eigenvalues, which are a key ingredient
to model the cosmic ionization by PBHs presented in
Paper~II. \cite{Tsuribe:95} has also investigated a spherically
symmetric accretion flow with an external radiation drag onto black
holes but for masses that are larger ($> 10^{4-5}$~M$_\odot$) than the
ones discussed in this paper. We will show in \S~\ref{ssec:selfs} that
in this mass range the steady-state assumption that defines the
``Bondi problem'' fails and the solutions transition to a class of
time-dependent self-similar solutions.

Another interesting mechanism that may lead to the formation in the
early universe of massive black holes has been proposed by
\cite{Loeb:93, Umemura:93}. These works discuss in detail the effect
of CMBR in the accretion process during a quasi-spherical collapse of
a self-gravitating gas clouds and the consequent formation of massive
black hole in the early universe, after recombination.  In this
scenario the removal of angular momentum due to Compton drag in
collapsing rare density perturbations at redshift $z \sim 400-1000$
leads to the formation of massive black holes with masses $M \simgt
10^5$ M$_\odot$. Radiation feedback effects associated with the
accretion process are also considered. These works differ from the
present one in several ways. They focus on the formation of black
holes rather than accretion onto a pre-existing black holes and the
typical masses of the black holes formed by this process are much more
massive than the ones considered in the present study. The accretion
is not steady and no Bondi solution exists. Although the black holes
formed in this scenario should be very rare, \cite{Sasaki:96} have
estimated that accretion onto such black holes would produce enough
radiation to reionize the universe by $z \sim 150$.

Several authors have investigated the effect that Compton drag and
Compton heating has on gas accretion in many different contexts
\citep[\eg,][]{FukeU:94, Mineshige:98, Ciotti:04, Park:07}.  Recently,
\cite{Wang:06} presented a discussion of feedback in different
accretion regimes, including a Bondi sphere in the case of a black
hole accretion at high redshift. They point out that Compton heating
can rapidly halt supercritical accretion onto remnant black holes of
$10^3$ M$_odot$ from Population~III stars. Such feedback would prevent
seed black holes to grow to masses typical of supermassive black holes
on timescale shorter than a Gyr. Thus, they conclude that accretion
onto seed black holes from population~III stars is not a viable
mechanism to explain the existence of quasars at redshift $z \sim 6$.
In the present paper the accretion onto PBHs is highly subcritical and
Compton heating is unlikely important. However, feedback processes and
other physical processes will be addressed in detail in Paper~II.

The organization of this paper is as follows. In \S~\ref{sec:eq} we
introduce the basic equations for accretion around a point mass in
comoving coordinates. In \S~\ref{sec:res} we solve numerically the
accretion equations and provide analytical fits for the accretion
eigenvalues. In \S~\ref{sec:dm} we repeat the calculations assuming
the gravitational potential of a dark halo with power-law density
profile. A summary and discussion is presented in \S~\ref{sec:sum}.

\section{Basic equations}\label{sec:eq}

At high redshift, repeated Compton scattering between free electrons
and CMB photons - assumed here to be homogeneous and isotropic -
act on the proper velocity of the gas, $v$, as a viscous term. The
forces exerted by the CMB photons are opposed to the motion of the
fluid element with respect to the photons' rest frame, producing a
mean restoring acceleration \citep{Peebles:80}
\[
{dv \over dt}=-{4 \over 3}{x_e \sigma_T U_{cmb} \over m_p
  c}v=-\beta v,
\]
where $x_e$ is the electron fraction, $\sigma_T$ is the Thompson cross
section to electron scattering and $U_{cmb}=aT_{cmb}^4$ is the CMB
energy density. By expressing the CMB temperature as a function of the
redshift, $z$, we have $\beta=2.06 \times 10^{-23}x_e(1+z)^4$
s$^{-1}$. Note that the typical time scale for Compton drag is $m_p/m_e
\sim 1000$ times longer than the timescale for Compton
cooling/heating.

The ``Bondi'' problem is described by the stationary equations of mass
and momentum conservation (including the pressure, gravitational
potential and viscosity term) and the equation of state of the gas.
In spherical coordinates the equations are:
\begin{equation}
\left\{
\begin{array}{cl}
\dot M_g &= 4 \pi r^2 \rho v,\\
v {dv \over dr} &= -{1 \over \rho}{dP \over dr} -{GM(<r) \over r^2} -
\beta(z) v,\\
P &= K \rho^\gamma.
\end{array}
\right.
\label{eq:1}
\end{equation}
Here, $v$, $r$ and $\rho$ are the velocity, distance from the center
of the gravitational potential and gas density. The ``dot'' represents
the time derivative, thus $\dot M$ is the mass accretion rate. The gas
pressure, $P$ is assumed to be a power law of the density with
exponent $1 \le \gamma\le 5/3$. This equation, know as the polytropic
equation of state, substitutes the energy conservation equation. The
case $\gamma=1$ corresponds to an isothermal system in which the
cooling time is much shorter than the dynamical time and the opposite
limit, an adiabatic system, is obtained for $\gamma=5/3$.

So far we have neglected the cosmological terms due to the Hubble
expansion and dark matter. Also, we have implicitly assumed that the
redshift-dependent viscous term, $\beta(z)$, changes slowly with
respect to the time scale necessary to achieve stationarity:
$t_{cr} \ll t_H$, where $t_{cr}$ is approximately the sound crossing
time and $t_H$ is the Hubble time. We will return to this point in the
next section.

\subsection{Hubble expansion: transition to self-similar  solutions}\label{ssec:selfs}

In order to include the cosmological term due to the Hubble expansion
we use the standard procedure of writing the hydrodynamic equations
(\ref{eq:1}) in a comoving frame of reference. This is done by
expressing the radial coordinate, $r$, and the radial velocity, $v$, in
terms of the comoving radius, $x$, and peculiar velocity, $v_p$:
\begin{equation}
\left\{
\begin{array}{cl}
r &= a(t)x,\\
v &={d \over dt}(a x)= Hr+v_p.
\end{array}
\right.
\label{eq:1b}
\end{equation}
Here, $a(t)$ is the scale factor, $H=\dot a/a$ is the Hubble parameter
and the peculiar velocity is defined as $v_p=a {\dot x}$. The
stationary ($\partial \delta /\partial t=0$, $\partial v_p /\partial
t=0$) equations for the overdensity, $\delta=\rho/\overline \rho$, and
peculiar velocity, $v_p$, are
\begin{equation}
\left\{
\begin{array}{cl}
\dot M_p &= 4 \pi x^2 (1+\delta) v_p,\\
{v_p \over a}{\partial v_p \over \partial x} &= -{1 \over \rho}{dP
  \over adx} -{G[\Delta M(<ax)] \over
  (ax)^2} - (\beta + H) v_p
\end{array}
\right.
\label{eq:2}
\end{equation}
Multiplying both sides of the mass conservation equation (top equation
in [\ref{eq:2}]) by $\overline \rho(t) a(t)^2$, we recover the same
equation as in (\ref{eq:1}), where $\dot M_g= \overline \rho(t) a(t)^2
\dot M_p$ is now a function of time (or redshift).  We also notice
that $\partial /\partial r = a^{-1}\partial /\partial x$ and that
$\Delta M(<r) = M(<r) - 4\pi \overline \rho r^3/3 \approx
M(<r)$. Thus, neglecting the {\it self-gravity of the accreted gas},
the momentum conservation equation (the bottom equation in
[\ref{eq:2}]) has the same form as in (\ref{eq:1}) after replacing
$\beta$ with the effective viscosity $\beta^{eff}=(\beta+H)$:
\begin{equation}
\left\{
\begin{array}{cl}
\dot M_g &= 4 \pi r^2 \rho(z) v_s,\\
v_s {dv_s \over dr} &= -{1 \over \rho}{dP \over dr} -{GM(<r) \over r^2} -
\beta^{eff}(z) v_s,\\
P &= K \rho^\gamma.
\end{array}
\right.
\label{eq:3}
\end{equation}
If we neglect the gas self-gravity, the cosmological expansion does not
change the basic form of the equations but introduces: (i) an
additional viscosity term in the momentum equation and (ii) the boundary
conditions at $r \rightarrow \infty$ for the density $\rho_\infty(z)$
and the sound speed $c_{s,\infty}(z)$ are a function of redshift.
Although we are assuming stationarity in the comoving frame of
reference, the flow is time-dependent in physical coordinates. The
assumption of steady flow is justified as long as the time scale to
achieve the stationary solution in the comoving frame of reference is
shorter than the Hubble time.  We have already made this assumption
when we have considered the Compton viscosity, $\beta(z)$.

\subsubsection{Stationarity}

The typical scale of the system is the Bondi radius $r_b$, hence, the sound
crossing time is $t_{cr} \sim r_b/c_{s, \infty}$. If $t_{cr}< t_H$ the
Bondi solution is valid because the cosmological terms vary slowly
with respect to the typical time scale of the system and a
quasi-steady solution can be found.

The analysis of the dimensionless accretion equations (see
\S~\ref{sec:res}) show that the viscosity term due to the Hubble
expansion becomes important when $H r_b/c_{s, \infty} > 1$. This
condition is equivalent to the constrain $t_{cr} > t_H$.  Thus, the
assumption of stationarity, implicit in the derivation of the Bondi
solution, fails when the Hubble term becomes important. In
\S~\ref{sec:sum}, assuming a static PBH and neglecting feedback
effects, we estimate that the hypothesis of stationarity fails for PBH
masses $M_{PBH} \simgt 2 \times 10^4$~M$_\odot$. This is a
conservative estimate because if we consider the proper motion of PBHs
and feedback effects we derive a larger critical mass.  However, the
existence of PBHs more massive than $10^4-10^5$ M$_\odot$ is unlikely,
so the cases for which the Bondi solution does not apply are not
particularly relevant for the aim of the present study.

\begin{figure*}[ht]
\epsscale{1.0}
\plottwo{\figname{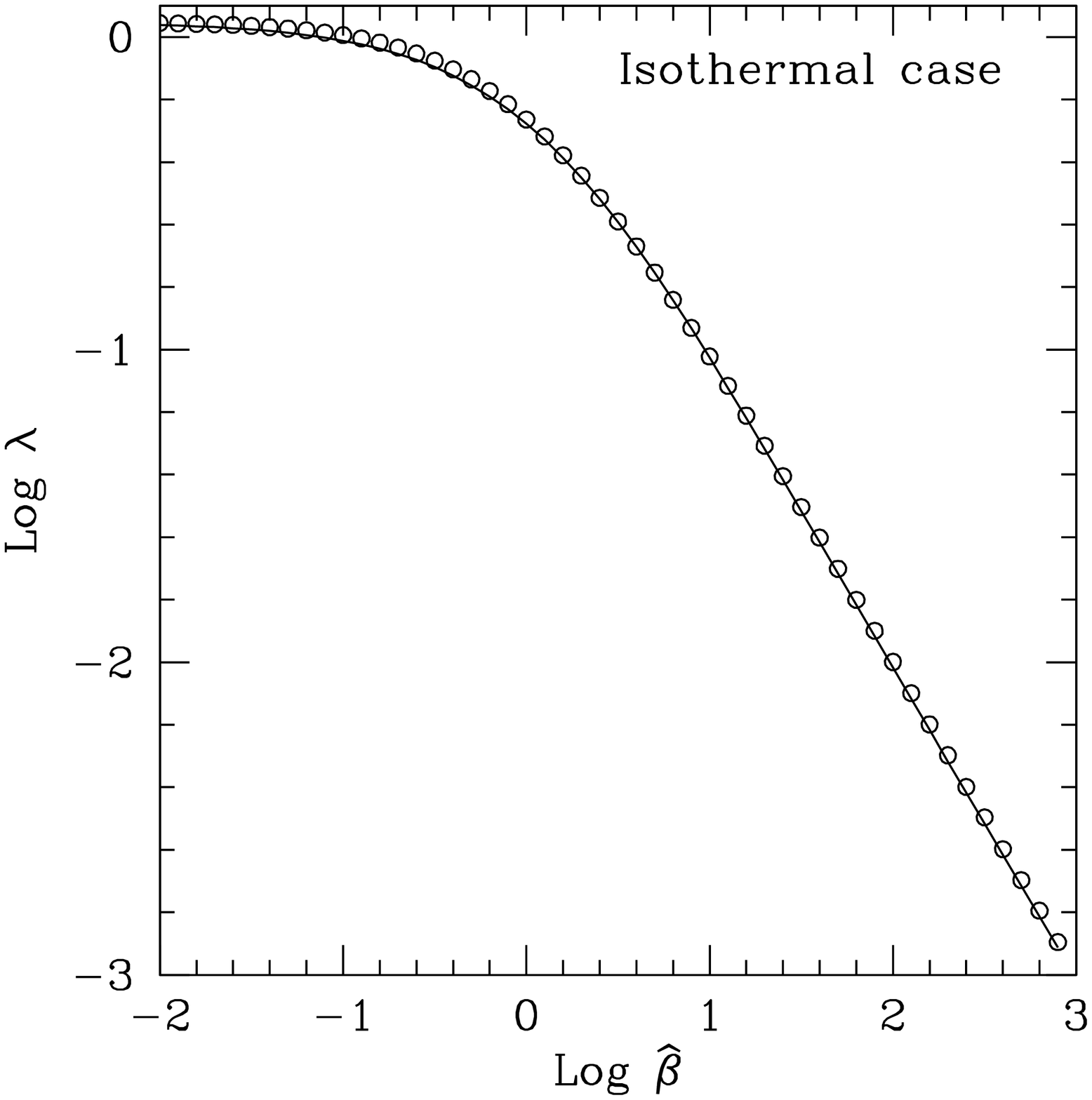}}{\figname{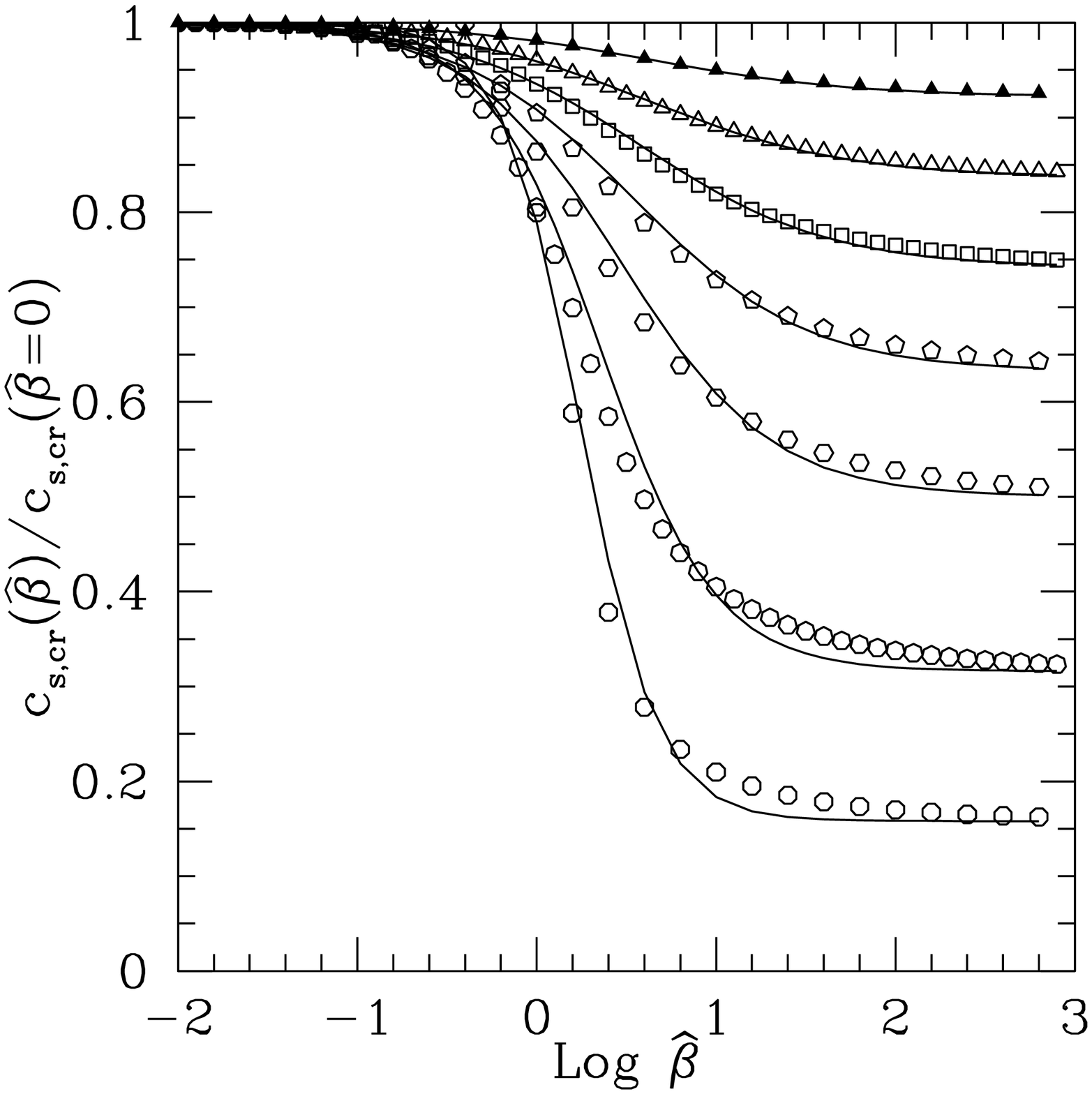}}
\caption{({\it Left}). Dimensionless accretion rate $\lambda$ onto a
  point mass as a function of the effective viscosity, $\hat \beta$,
  produced by Compton drag and Hubble expansion. We assume an
  isothermal equation of state for the gas. The points show the result
  of numerical integration and the curve shows the fit with the
  relationship in \protect{\eq~(\ref{eq:d_iso})}. ({\it Right}). The
  same as in the left panel but assuming a polytropic equation of
  state for the gas with $1<\gamma \le 5/3$. Instead of the accretion
  eigenvalue $\lambda$, here we show the sound speed at the sonic
  radius, $c_{s,cr}$ (normalized to its value neglecting viscosity,
  $\hat \beta=0$), as a function of the effective viscosity $\hat
  \beta$. The accretion rate can be calculated from
  \protect{\eq~(\ref{eq:lambda})}. The curves from top to bottom refer
  to $\gamma=1.1, 1.2, 1.3, 1.4, 1.5, 1.6$ and $1.65$, respectively.}
\label{fig:d_iso}
\end{figure*}

\subsubsection{Gas Self-gravity}

The assumption of neglecting the self-gravity of the gas is justified
if $M \gg M_{gas}$, where $M_{gas}$ is the total gas mass contained in
the halo around the black hole. It is easy to estimate the gas mass,
$M_{gas}^B$, in the limit in which the viscosity terms are negligible
(\ie, for the classic ``Bondi'' solution). This estimate is a lower
limit for $M$ because $M \gg M_{gas}^B > M_{gas}$.  By integrating the
density profile for the Bondi solution we find that $M_{gas}^B \sim
8\pi/3 \rho_\infty r_b^3$, where $r_b=GM/c^2_{s,\infty}$ is the Bondi
radius. Thus, for $M \ll M^{cr}=(8\pi
G^3/3)^{-1/2}c_{s,\infty}^3\rho_\infty^{-1/2}$ we can neglect the gas
self-gravity.  In the case of PBHs accreting from the IGM between the
redshift of recombination and decoupling ($100 \simlt z \simlt 1000$)
we have $c_{s,\infty} \approx 5.74 {\rm km~s^{-1}} [(1+z)/1000]^{1/2}$
and $\rho_\infty = \rho_{cr} \Omega_b (1+z)^3$. Assuming $\Omega_bh^2
=0.02$ we find $M^{cr} \sim 3 \times 10^{5}$~M$_\odot$.  At later
times ($30 \simlt z \simlt 100$) the critical PBH mass may become
larger than $M^{cr}$ because the temperature of the IGM may increase
due to the radiation emitted by the gas accreting on the PBHs. The
condition for steady accretion is more restrictive than the one found
here. Thus, if steady solutions exist, the gas self-gravity can always
be neglected.

The solution of the Bondi problem including the gas self-gravity has
been studied in the context of stellar structure. The solution is
known to be unstable for $\gamma < 4/3$.  In this limit we would find
that all the gas in the universe is eventually accreted onto PBHs. In
the regime discussed here only a small fraction of the cosmic gas is
accreted.

\subsubsection{Dark matter self-gravity}

The assumption of neglecting self-gravity of the accreting material is
a good approximation for the baryons but not for the collisionless
dark matter component.  Unless all dark matter is constituted of
primordial black holes (\ie, $\Omega_{dm}=\Omega_{bh}$) we need to
take into account the growth of structures seeded by the presence of
PBHs \citep{Mack:06}.  The total energy of a representative patch of
the Universe at critical density is zero. Thus, the volume is
marginally stable and the introduction of any perturbation, such as a
PBH, produces the collapse of the region.

\cite{Mack:06} found that a dark matter halo about 100 times more
massive than the primordial black hole at its center is accreted by
redshift $z\sim 30$.  The accretion takes place mostly after the epoch
of ``matter-radiation'' equivalence.  Calculations of the gas
accretion rate into the gravitational potential produced by the dark
matter halo will be considered in \S~\ref{sec:dm}. When a dark
halo grows, the gas self-gravity can be neglected with respect to the
dark matter potential because the total gas mass is about five times
smaller (\ie, the cosmic value) than the dark matter.

\section{Spherical accretion onto a point mass}\label{sec:res}

\subsection{Sonic point}

It is convenient to write \eq~(\ref{eq:3}) in dimensionless units,
$x=r/r_b$, ${\cal M}=v_s/c_s$, $\hat c_s=c_s/c_{s,\infty}$, $\hat
\rho=\rho/\rho_\infty$, where $c_{s,\infty}$ and $\rho_\infty$ are the
sound speed and gas density at radii $r \rightarrow \infty$,
respectively. The scaling constant for the radius, viscosity and
accretion rate are:
\begin{equation}
r_b ={GM_{bh} \over c_{s,\infty}^2}, ~~~ \hat \beta =\beta^{eff} {r_b \over c_{s,\infty}}
~~~ \lambda = {\dot M_g \over 4 \pi r_b^2 \rho_\infty c_{s,\infty}}.
\end{equation}
We obtain the familiar set of dimensionless equations:
\begin{equation}
\left\{
\begin{array}{cl}
\lambda &= \hat \rho^{\gamma+1 \over 2} {\cal M}x^2\\
{\dot {\cal M} \over {\cal M}} &= {{2 \over x}(1+{(\gamma-1) \over 2}{\cal M}^2)-{\gamma+1 \over 2}\left({1 \over \hat c_s^2 x^2}+{\hat \beta
  {\cal M} \over \hat c_s}\right) \over {\cal M}^2-1}\\\label{eq:drag2}
\hat c_s &= \hat \rho^{\gamma-1 \over 2}
\end{array}
\right.
\end{equation}
The transonic solution crosses the sonic point (${\cal M}=-1$) at
the critical radius:
\begin{equation}
x_{cr}=\hat c_{s,cr}^{-2}{-1+\sqrt{1+\hat \beta\hat c_{s,cr}^{-3}} \over
  \hat \beta \hat c_{s,cr}^{-3}},
\label{eq:sonic}
\end{equation}
where $\hat c_{s,cr}$ is the dimensionless sound speed at the critical
radius.  In the limit of small effective viscosity (\ie, $\hat \beta
\hat \ll c_{s,cr}^{3}$) we find $x_{cr} \rightarrow (2 \hat
c_{s,cr}^2)^{-1}$. In the limit of large effective viscosity we find
$x_{cr} \rightarrow (\hat c_{s,c}\hat \beta)^{-1/2}$.  The eigenvalue,
$\lambda$, for the transonic solution is
\begin{equation}
\lambda =\hat \rho_{cr}^{\gamma+1 \over 2} x_{cr}^2=\hat c_{s,cr}^{\gamma+1 \over \gamma-1}x_{cr}^2
\label{eq:lambda}
\end{equation}
where $\hat \rho_{cr}$ is the dimensionless density at the critical radius.

\subsection{Accretion eigenvalues}

\subsubsection{Isothermal equation of state}
 
Since we did not find analytic solutions for \eq~(\ref{eq:drag2}), we
integrated the equations numerically. We found the asymptotic
solutions of the equations in the limits of negligible viscosity,
$\hat \beta \rightarrow 0$, and large viscosity, $\hat \beta
\rightarrow \infty$. In the limit of small viscosity we find the
classic Bondi solution $\lambda \rightarrow \exp({3/2})x_{cr}^2$,
where $x_{cr}=0.5$. In the limit of large effective viscosity, the
asymptotic analysis of \eq~(\ref{eq:drag2}) shows that $\hat \rho_{cr}
\rightarrow 1$ and that $\lambda \rightarrow \hat{\beta}^{-1}$. We
have used the asymptotic solutions for $\lambda$ to guess the form of
the appropriate fitting function of the numerical results. The
function
\begin{equation}
\lambda =\exp\left[{9/2 \over 3+\hat{\beta}^{~0.75}}\right]x_{cr}^2,
\label{eq:d_iso}
\end{equation}
gives a good fit for the numerical results as shown in \fig~\ref{fig:d_iso}(left).

\subsubsection{Polytropic equation of state}

The asymptotic solutions for the cases of negligible effective
viscosity and large effective viscosity can easily be found.
Neglecting the viscosity terms we have $\hat{c}_{s, cr} \rightarrow
[2/(5-3\gamma)]^{1/2}$. In the limit of large viscosity the asymptotic
analysis of \eq~(\ref{eq:drag2}) shows that $\hat \rho_{cr}
\rightarrow 1$, $\hat{c}_{s, cr} \rightarrow 1$ and $\lambda
\rightarrow \hat \beta^{-1}$. We solve \eq~(\ref{eq:drag2})
numerically and find a parametric fit for the numerical results, with
the correct asymptotic behavior for the sound speed at the critical
radius. We also make sure that in the limit $\gamma \rightarrow 1$,
the parametric fit for $\hat c_{s, cr}$ and $\lambda$ have the same
expression as that found for the isothermal case. We find
\begin{eqnarray}
\hat c_{s,cr}(\hat \beta) &=&{3\hat c_{s,cr}(0)+\hat{\beta}^\xi \over 3+\hat{\beta}^\xi},\nonumber\\
\xi &=& 0.75\left({\gamma+1 \over 5-3\gamma}\right)^{\gamma-1 \over \gamma+1},
\label{eq:d_ad}
\end{eqnarray}
where $\hat c_{s, cr}(0) = [2/(5-3\gamma)]^{1/2}$ is the sound speed
at the critical radius for $\hat \beta=0$. The fitting curves (solid
lines) and the numerical integrations (shown by the symbols) are shown
in \fig~\ref{fig:d_iso}(right). The eigenvalue $\lambda$ can be
calculated for any given value of the polytropic index, $\gamma$,
combining equations~(\ref{eq:lambda}) and (\ref{eq:d_ad}).

\section{Spherical accretion into a dark matter halo}\label{sec:dm}

If dark matter is entirely composed of PBHs each of them would
only accrete gas. In all the other cases a dark halo of WIMPs forms
around each PBH.  In this section we examine the gas accretion rate
into the gravitational potential of a spherical dark matter halo with
power-law density profile. The assumption of a single power-law for the density
profile is partially dictated by the need for simplicity but it is
also justified theoretically as elucidated below.
\begin{figure*}[ht]
\epsscale{1}
\plottwo{\figname{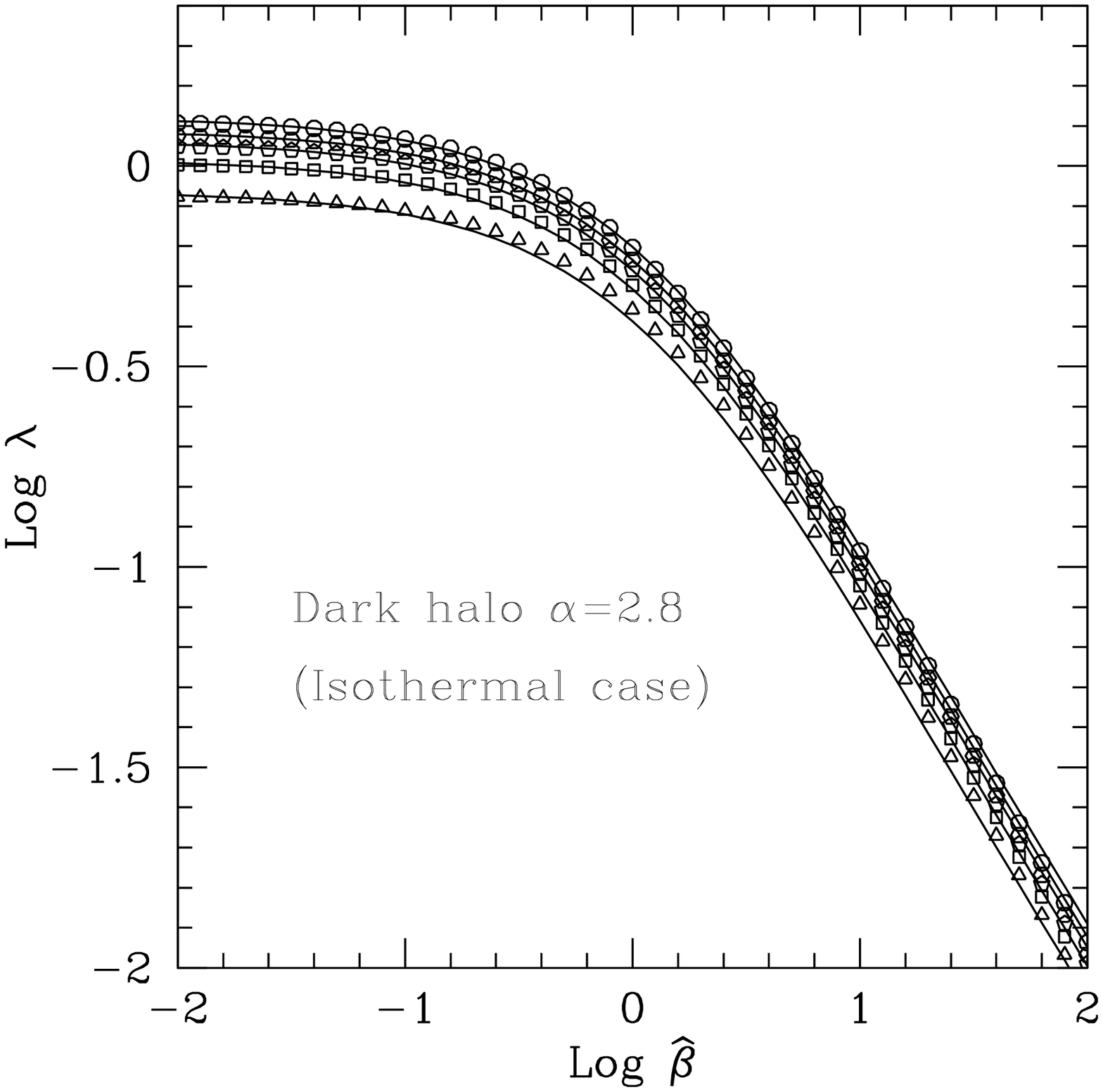}}{\figname{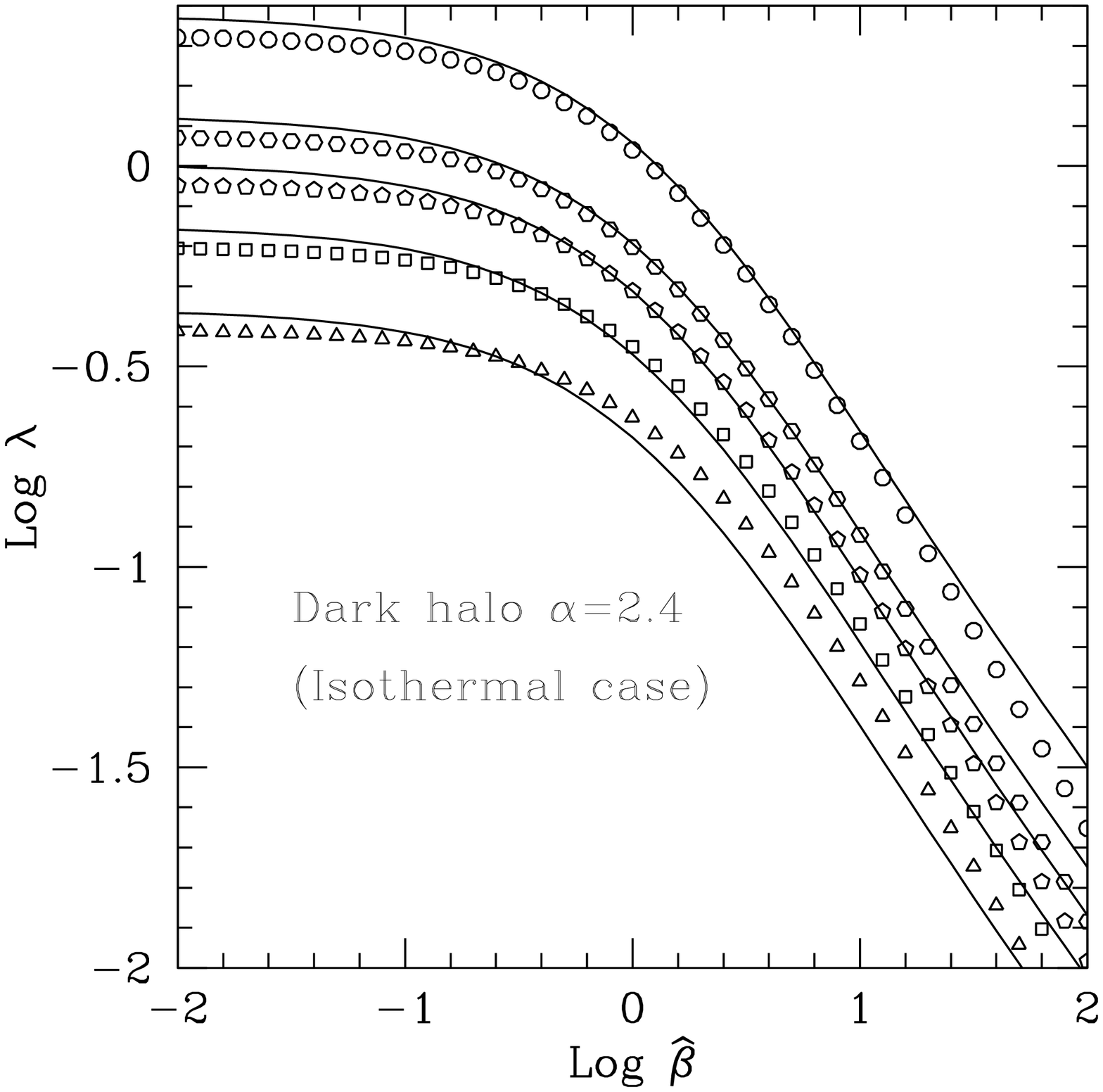}}
\caption{({\it Left}). Dimensionless accretion rate $\lambda$ into a
  spherical dark matter halo as a function of the effective viscosity,
  $\hat \beta$, produced by Compton drag and Hubble expansion. Each
  curve, from bottom to top, refers to halos of increasing radius. If
  the radius of the halo is smaller than the accretion radius, $r_B$,
  defined in the text, the accretion rate into an extended halo equals
  the one calculated for a point mass. We assume isothermal equation
  of state for the gas and a power-law density profile of the dark
  halo with slope $\alpha=2.8$. From bottom to top the symbols show
  the result of numerical integration and the curves show the fits
  with the parametric relationship in
  \protect{\eq~(\ref{eq:d_iso_halo})} for values of the dimensionless
  halo radius $r_B/r_{h}=\chi^{1/(1-p)}=2, 1, 0.5, 0.25$ and
  $0.016$. ({\it Right}). The same but for $\alpha=2.4$.}
\label{fig:d_iso_halo}
\end{figure*}

PBHs accumulate a spherical halo of dark matter that grows in mass
proportionally to $(1+z)^{-1}$. Non-negligible growth of the dark
matter halo begins at matter-radiation equality, and by redshift $z
\simeq 30$ the black hole is embedded in a halo about 100 times its
original mass \citep{Mack:06}.  The numerical values of the halo mass
and the comoving turnaround radius are:
\begin{eqnarray}
M_h &=& 3 M_{PBH} \left({1+z \over 1000}\right)^{-1},\\ 
r_{ta,com} &=& 58~{\rm pc}\left({M_{h} \over 1~M_\odot}\right)^{1/3}.
\end{eqnarray}

The theory of secondary infall predicts that the dark halo has a
radial density profile that is well approximated by a single power-law
cusp and a truncation of the density profile at a fraction of the
turnaround radius. All of the halo mass is contained within the cusp.
Two types of time-dependent, self-similar solutions have been found
for the dark matter profile \citep{GunnG:72, FillmoreG:84,
Bertschinger:85}; the transition from one type to the other depends on
the amount of angular momentum of the infalling matter:

\noindent
a) If the flow is highly radial such that the PBH at the center
absorbs all of the infalling matter, then almost all the mass in the
accreting halo is eventually incorporated into the central black hole.
This infall solution is characterized by the free-fall boundary
condition at the center and pressureless matter at infinity. The
solution is also valid for ``cold'' gas accretion because there is no
shell crossing and no standing shock forms.  The self-similarity of
the solution is valid if $M_{h} > M_{PBH}$, when the matter is
self-gravitating.  Near the center the halo mass is negligible
compared to the black hole mass, so the matter is free falling and the
density profile is $\rho \propto r^{-3/2}$. This slope is the same as
for the Bondi solution inside the sonic radius. In the outer parts the
solution is very different from the Bondi solution. The radial density
profile, due to the Hubble expansion, decreases with radius
propotionally to $r^{-3}$. The radius $r_h$ of the ``free-fall'' cusp
depends on the details of the collapse but is typically $r_h/r_{ta}\ll
1$ \citep{Mack:06}.
 
\noindent
b) If the flow is quasi spherical but the angular momentum of the
accreting dark matter prevents direct accretion onto the central black
hole, the particles pass near the center without being absorbed. For
the case in which the accreting matter is collisional, a standing
shock inside the turnaround radius stops the gas infall within the
core.  Remarkably, the solution for collisionless matter is very
similar to the $\gamma=5/3$ solution for gas accretion.  The density
profile has slope $\alpha=2.25$ at $r< r_h =0.339r_{ta}$. For $r>r_h$
the overdensity is close to one.

Estimates of the angular momentum of the accreting dark matter onto
PBHs are presented in a companion paper \citep{RicottiOM:06}. Based on
the results of the angular momentum calculations case (b) is the most
relevant. Nevertheless, if a fraction of the dark matter is captured
by the central PBH, we expect that the density profile has properties
intermediate between case (a) and (b). The truncation radius $r_h$
would be smaller than $r_{ta}/3$, and the effective slope of the inner
density profile steeper than $2.25$, in order to take into account the
non-negligible mass of the central black hole. Indeed a halo with
slope of the density profile $\alpha=3$ has a potential that
approaches the one of a point mass with $M \sim M_h$. For the sake of
generality we take this effects into account by treating the core
radius $r_h$ and slope of the core density profile as free parameters.

We calculate of the gas accretion rate by replacing the gravitational
acceleration for a point mass ($g_{bh}=-GM_{bh}/r^2$) in
\eq~(\ref{eq:drag2}) with that for an extended dark halo with
power-law density profile: $\rho=\Delta \rho_{dm}(z)
[r/r_{h}]^{-\alpha}$. Here, $\rho_{dm}(z)$ is the mean dark matter
density at redshift z and $\Delta$ is the overdensity of the halo at
$r=r_h$. We assume that all the mass of the halo is within the radius
$r_h$.  Thus, the gravitational acceleration is
\begin{equation}
g_{dm}(r) = -{GM_{h} \over r^2}{\rm min}[(r/r_{h})^p,1],
\label{eq:g_dm}
\end{equation}
where $p=3-\alpha$. Note that the halo radius is smaller than the
turnaround radius: $r_h<r_{ta}$.  We explore the range of values $2 <
\alpha < 3$ for the log-slope of the density profile (\ie, $0 < p <
1$).  If $p \rightarrow 0$ the gravitational acceleration approaches
the point-mass value. If the halo radius is $r_{h} \le x_{cr}= 0.5
r_b=0.5 GM_{h}/c_{s,\infty}^2$, the accretion eigenvalue is given by
\eq~(\ref{eq:d_iso}), the same as replacing the halo with a point-mass
$M=M_{h}$.  Thus, it is apparent that the accretion rate onto an
extended halo is a function of the dimensionless parameter
$\chi=r_b/r_{h}$ and the parameter $p$.  The case $p=1$ describes the
density profile for an isothermal sphere for which the circular
velocity is a constant. This case is singular because the Bondi
radius, for which $v_{cir}(r_b)=c_s$, cannot be defined.

\subsection{Sonic point}

After substituting $g(r)$ in \eq~(\ref{eq:3}) with the expression
given in \eq~(\ref{eq:g_dm}) for $g_{dm}(r)$, we rewrite the equations
in dimensionless units, $x=r/r_B$, ${\cal M}=v_s/c_s$, $\hat
c_s=c_s/c_{s,\infty}$, $\hat \rho=\rho/\rho_\infty$, where
$c_{s,\infty}$ and $\rho_\infty$ are the sound speed and gas density
respectively, at radii $r \rightarrow \infty$. Note that the scale
radius, $r_B$, for spherical accretion onto an extended halo of mass
$M_{h}$ is not defined analogously to point-mass Bondi radius,
$r_b=GM_{h}/c_{s,\infty}^2$, of equivalent total mass. It is
convenient to express the dimensional constants for an extended halo
as a function of the parameter $\chi = r_b/r_{h}$:
\begin{equation}
r_B =r_{h}\chi^{1 \over 1-p}=r_{b}\chi^{p \over 1-p}
\end{equation}
where $1 \le (1-p)^{-1} \le \infty$ and $0 \le p/(1-p) \le \infty$ for
$2 < \alpha < 3$. The dimensionless viscosity and accretion rate are
defined as:
\begin{equation}
\hat \beta =\beta^{eff} {r_b \over c_{s,\infty}} \chi^{p \over 1-p}, ~~~ \lambda = {\dot M_g \over 4 \pi r_b^2 \rho_\infty c_{s,\infty}}\chi^{-{2p \over 1-p}}.
\end{equation}
In these units we obtain the dimensionless equation:
\begin{equation}
{\dot {\cal M} \over {\cal M}} = {{2 \over x}(1+{(\gamma-1) \over 2}{\cal M}^2)-{\gamma+1 \over 2}\left({1 \over \hat c_s^2 x^{2-p}}+{\hat \beta
  {\cal M} \over \hat c_s}\right) \over {\cal M}^2-1}.
\end{equation}
The transonic solution crosses the sonic point (${\cal M}=-1$) at
the critical radius $x_{cr, dm}$, found by solving the equation
\begin{equation}
\hat \beta \hat c_{s,cr}x_{cr,dm}^{2-p}+2 \hat c_{s,cr}^{2}x_{cr,dm}^{1-p}-1=0,
\label{eq:sonic1}
\end{equation}
where $\hat c_{s,cr}$ is the dimensionless sound speed at the critical
radius. It is useful to find the asymptotic behaviors of the critical
radius in the limit of zero viscosity, $x_{cr, dm} \rightarrow (2\hat
c_{s,cr}^{2})^{-1/(1-p)}$ and in the limit of large viscosity $x_{cr,
dm} \rightarrow (\hat c_{s,cr}\hat \beta)^{-1/(2-p)}$.  These
asymptotic behaviors will guide us to find the functional form of the
fitting formula.

\subsection{Accretion eigenvalues}

For the sake of simplicity, given that we have introduced two new free
parameters that describe the halo potential, we will only discuss the
case of an isothermal equation of state for the gas. Physically, the
isothermal equation of state is the most relevant case because the CMB
photons that exert Compton drag are about 1000 times more efficient in
keeping the gas temperature constant, and close to the CMB value.

\subsubsection{Isothermal equation of state}

The accretion eigenvalue can be fitted with the following approximate formula
\begin{equation}
\lambda^{h} =f_{\beta} f_{\chi}\exp\left[{9/2 \over 3+\hat{\beta}^{~0.75}}\right]x_{cr,dm}^2,
\label{eq:d_iso_halo}
\end{equation}
where the functions
\begin{eqnarray}
f_\beta &=& 1+1.25\hat{\beta}^{p \over 2-p},\\
f_\chi&=&\left[\exp(2-\chi)\right]^{p \over 1-p},
\end{eqnarray}
are small corrections to $\lambda$ that vanish for small values of $p$
and $\chi \rightarrow 2$. This formula cannot be used without an
expression for $x_{cr,dm}$.  The function $f_\beta$ has been derived
from the asymptotic analysis of the equation for the critical radius
$x_{cr,dm}$.  Assuming an isothermal equation of state, the sonic
radius approaches $x_{cr,dm} \rightarrow 2^{-1/(1-p)}$ in the limit of
negligible effective viscosity, and $x_{cr,dm} \rightarrow \hat
\beta^{-1/(2-p)}$ in the limit of large effective viscosity. Thus, the
sonic radius of the transonic solution for accretion into a halo
potential is related to that of a point-mass (with $M=M_{h}$) by the
approximate fitting formula
\begin{equation}
{x_{cr,dm} \over x_{cr}}\approx 2^{-{p \over 1-p}}\left({g_\beta \over f_\beta}\right)^{1/2},
\label{eq:xcr_halo}
\end{equation}
where
\begin{equation}
g_\beta= \left(1+10\hat{\beta}\right)^{p \over 10(1-p)}.
\end{equation}

Combining \eqs~(\ref{eq:xcr_halo}) and (\ref{eq:d_iso_halo}) we can
relate the accretion eigenvalue for an extended dark matter halo to
the eigenvalue in \eq~(\ref{eq:d_iso}) for a point mass $M=M_{h}$:
\begin{equation}
{\lambda^{h} \over \lambda^{bh}}=2^{2p \over 1-p}g_{\beta}f_\chi.
\label{eq:d_iso_halo1}
\end{equation}
The fits using the parametric function \eq~(\ref{eq:d_iso_halo}) to
the numerical integrations are shown in \fig~\ref{fig:d_iso_halo} for
different values of $\alpha$ and the parameter $\chi$.

It is important to note that the dimensional units for the accretion
rate, radius and effective viscosity are different for the cases of a
dark matter potential and a point-mass. Thus the dimensional accretion
rate into a halo is reduced by a factor $\chi^{2p/(1-p)}$ and the
effective viscosity increases by a factor $\chi^{p/(1-p)}$ with
respect to the point-mass case. To summarize, the accretion rate into
a halo of mass $M_{h}$ is related to the accretion rate onto a black
hole of equivalent mass $M_{bh}$ by the relationship
\begin{eqnarray}
\hat \beta^{h} &\equiv& \chi^{p \over 1-p}\hat \beta^{bh},\\
\lambda^{h} &\equiv& \Upsilon^{p \over 1-p} \lambda^{bh},\\
r_{cr}^{h} &\equiv& x_{cr,dm}r_B= \left({g_\beta \over f_\beta}\right)^{1/2}\left({\chi \over 2}\right)^{p \over 1-p}r_{cr},
\end{eqnarray}
where
\begin{equation}
\Upsilon = \left(1+10\hat{\beta}^{h}\right)^{1 \over 10}
\exp{(2-\chi)} \left({\chi \over 2}\right)^2.
\end{equation}
Note that $\hat \beta^h$ and not $\hat \beta^{bh}$ should be used in the analytic
expressions for $\lambda^{bh}$ and $r_{cr}$.

\begin{figure*}[ht]
\epsscale{1}
\plottwo{\figname{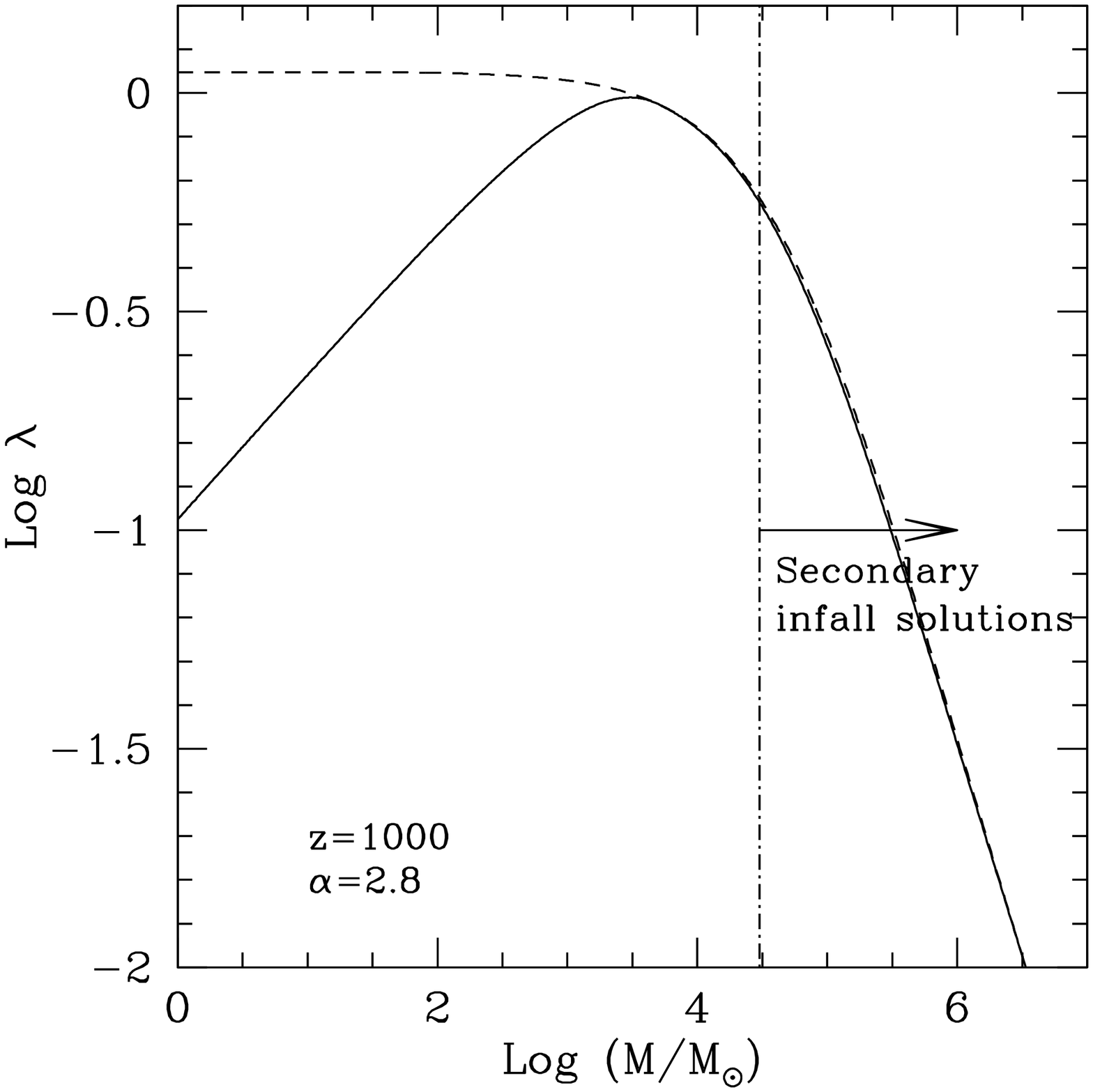}}{\figname{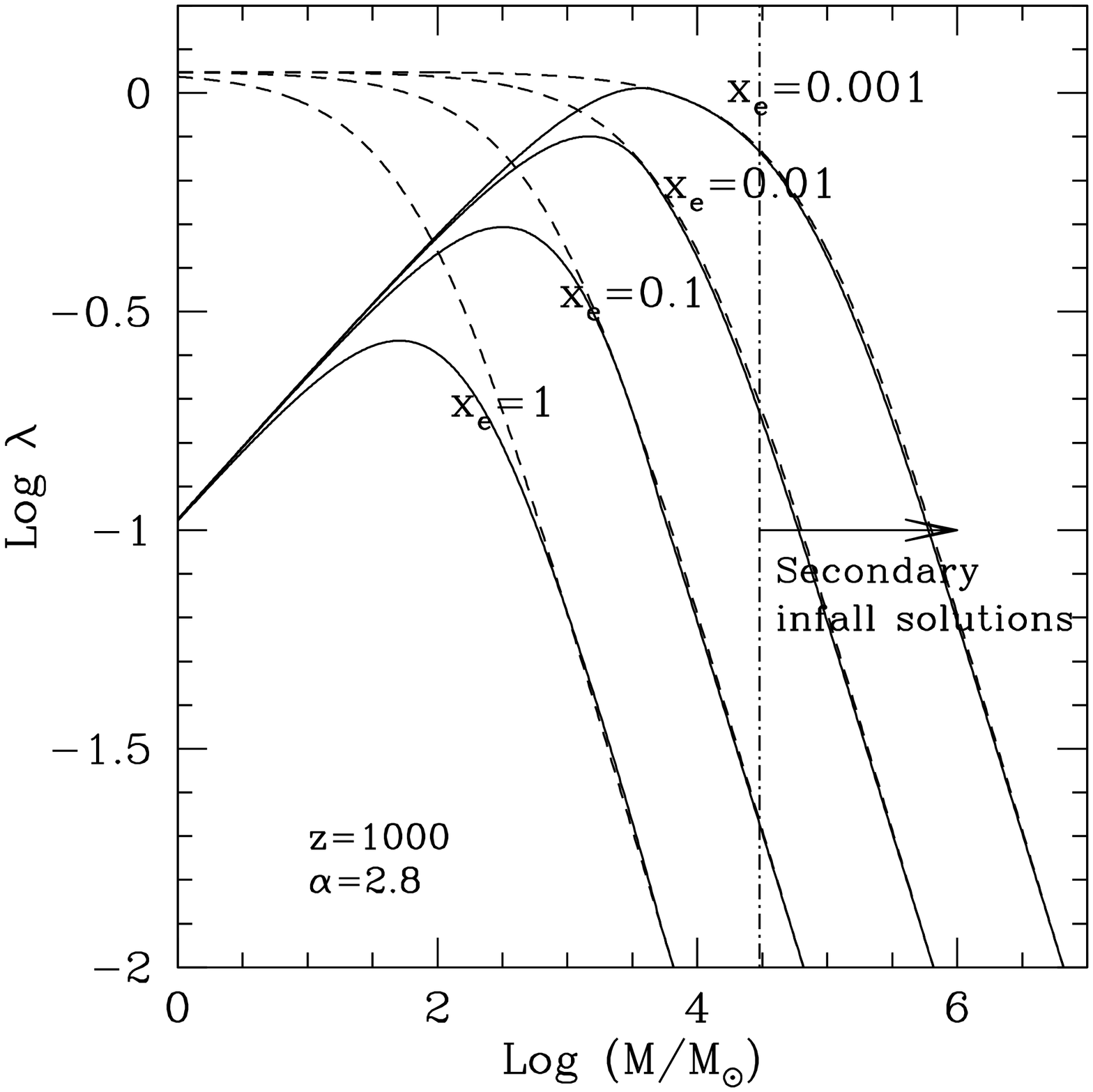}}
\caption{({\it Left}). Dimensionless accretion rate $\lambda$ into a
  spherical dark matter halo (solid curve) and a point mass (dashed
  curve) as a function of the mass of the object (PBH or dark halo) at
  $z+1=1000$. We have considered only the viscosity from the Hubble
  expansion and a halo with power-law slope $\alpha=2.8$.  ({\it
  Right}). The same as the left panel but considering only the
  viscosity due to Compton drag. The different curves refer to gas
  with fractional ionization $x_e=1, 0.1,0.01$ and $0.001$ as shown by
  the labels.}
\label{fig:t_masses}
\end{figure*}

\section{Summary and Discussion}\label{sec:sum}

This paper examines the classic Bondi problem of steady, spherical gas
accretion in a cosmological framework.  The main motivation of the
present work is to determine the accretion rate onto primordial black
holes in the early universe, before the formation of the first
galaxies and stars.  We estimate the importance of cosmological
effects such as the gas viscosity due to the coupling of the gas to
the CMB photon fluid (Compton drag), the Hubble expansion and the
formation of a dark halo around the black hole. In a companion paper
we use the results of the present work to model cosmic ionization
by the X-rays emitted from accreting PBHs. Comparisons between
theoretical models and WMAP3 data allow us to constrain mass
and abundances of PBHs \citep{RicottiOM:06}.

The cosmological Bondi problem is tightly related to the theory of
``secondary infall'' developed by \cite{GunnG:72, FillmoreG:84,
Bertschinger:85}.  In the standard cosmological model the baryons are
a small fraction of the dark matter, hence the evolution of dark
matter is unaffected by the presence of the gas.  We use the results
of the secondary infall theory to determine the shape of the
gravitational potential into which the gas is infalling.  In addition,
when the viscosity term due to cosmic expansion becomes important, the
assumption of steady flow for the gas, implicit in deriving the Bondi
solution, becomes invalid. In this limit, time-dependent infall
solutions should be used instead.  In \S~\ref{ssec:infall} we discuss
the relationships between the Bondi and the secondary infall solutions
and the conditions of validity for each of them.

\subsection{Accretion onto a point mass}\label{ssec:pm}

The physical values for the accretion rate and the critical radius are
easily calculated from the dimensionless ones. The values of the
Bondi-Hoyle radius, $r_b$, and accretion rate are
\begin{eqnarray}
r_b&\approx&43.3 ~{\rm pc} \left({M \over 10^{4} M_\odot}\right)
\left({ c_{s,\infty} \over 1 {\rm km~s^{-1}}}\right)^{-2},\\ 
\dot M &\approx&\lambda (3.7 \times 10^{22}~{\rm g~s^{-1}})n_{\rm gas}
\left({M \over 10^{4} M_\odot}\right)^2 \left({ c_{s,\infty} \over 1
{\rm km~s^{-1}}}\right)^{-3},
\label{eq:dim}
\end{eqnarray}
respectively \citep{BondiH:44}.  The accretion rate eigenvalue,
$\lambda$, is a function of the dimensionless ``effective viscosity'',
$\hat \beta=\beta^{eff}t_{cr}$, where the Bondi radius crossing time
is $t_{cr}=r_b/c_{s,\infty}$ and $\beta^{eff}=\beta(z)+ H(z)$.  In the
relevant redshift range (\ie, before $z\sim 30$) the Compton drag term
is $\beta(z) \propto x_e(1+z)^4$ and the Hubble parameter is $H(z) =
3.24 \times 10^{-18} s^{-1}(\Omega_m h^2)^{1/2}(1+z)^{3/2}$. It
follows, assuming $\Omega_m h^2=0.127$ \citep{Spergel:06}, that
$\beta^{eff}=H(z)\{1+1.78 x_e [(1+z)/100]^{5/2}\}$ and the Compton term
is dominant over the Hubble term at redshifts
\begin{equation}
z > 500 \left({x_e \over 0.01}\right)^{-2/5}.
\end{equation}
The gas viscosity can be neglected if $\beta^{eff}t_{cr} < 1$, where
\[
t_{cr}={r_b \over c_{s,\infty}}=(0.224~{\rm Myr})\left({M \over 10^{4}~M_\odot}\right) \left({
c_{s,\infty} \over 5.74 ~{\rm km~s^{-1}}}\right)^{-3}.
\]
Including the Hubble flow and Compton drag terms, the
expression for the dimensionless effective viscosity is 
\begin{eqnarray}
\hat \beta=&\left({M \over 10^4 M_\odot}\right)\left({z+1 \over 1000}\right)^{3/2}\left({c_{s,\infty} \over 5.74 {\rm km s^{-1}}}\right)^{-3}\nonumber\\
&\times\left[0.257+1.45\left({x_e \over 0.01}\right)\left({z+1 \over 1000}\right)^{5/2}\right].
\end{eqnarray}

Let us consider first the viscosity term due to the Hubble expansion.  We
can neglect the Hubble flow when $t_{cr}<t_H=H(z)^{-1}$ or,
equivalently, when $v_H/c_{s,\infty}<1$, where $v_H=r_b H(z)$ is the
typical Hubble velocity of the gas at the Bondi radius. If we assume
that the gas temperature equals the CMB temperature,
$c_{s,\infty}=1.1~{\rm km/s} (T_{cmb}/100 K)^{1/2}$, we find that
$t_{cr}<t_H$ for
\begin{equation}
M_{pbh} < 2.7 \times 10^4~{\rm M}_\odot.
\end{equation}
Hence, the accretion onto primordial black holes less massive than $3
\times 10^4$ M$_\odot$ (or more massive if the gas temperature is
$T_{gas}>T_{cmb}$ due to photo-heating from accreting PBHs) is weakly
affected by the Hubble expansion. But, as discussed in detail in
\S~\ref{ssec:infall}, when the Hubble viscosity becomes important the
hypothesis of quasi-steady flow fails and the Bondi solution
transitions to the time-dependent secondary infall solution.

Assuming that the gas temperature equals the CMB temperature, the
viscosity term due to Compton drag is negligible for
\[
M_{pbh} \simlt {70~{\rm M}_\odot \over x_e} \left({z+1 \over
    1000}\right)^{-5/2}.
\]
Before recombination, PBHs with mass $M_{pbh} \simlt 50-100$ M$_\odot$
are unaffected by Compton drag. Instead, the accretion rate onto PBHs
more massive than $100-1000$ M$_\odot$ is negligible before
recombination.  In the standard cosmological model the residual
ionization fraction after recombination is $x_e \sim 10^{-4}$, hence,
after recombination, Compton drag is negligible for any reasonable PBH
mass. Only if a sufficiently large number of PBHs or other sources of
ionizing radiation increase the fractional ionization above $x_e
\simgt 1.47 \times 10^{-3}$ the Compton viscosity becomes important
(\ie, dominant with respect to the viscosity term due to the Hubble
flow).

The eigenvalue $\lambda$ can be derived from the equation of mass
conservation (\ref{eq:lambda}) that depends on the polytropic index,
$\gamma$, the sonic point, $x_{\rm cr}$ [given by
\eq~(\ref{eq:sonic})], and the sound speed at the sonic point, $c_{\rm
s,cr}$.  We provide fits to the sound speed at the sonic point (see
\fig~\ref{fig:d_iso}) in \eq~(\ref{eq:d_ad}).  In the limit $\gamma=1$
(isothermal equation of state) the accretion eigenvalue is given by
\eq~(\ref{eq:d_iso}) and depends only on $x_{\rm cr}$.

\subsection{Accretion into a dark matter halo}\label{ssec:dm}

If the radius of the dark matter halo, $r_h$, is within the Bondi
critical radius, the accretion rate is unchanged with respect to the
case of a point mass. Quantitatively, the accretion rate is
reduced approximately by $(\chi/2)^{2p/(1-p)}$ with respect to the
point mass case where
\begin{equation}
{\chi \over 2}={r_{b} \over 2 r_{h}}=3.3 \times 10^{-3}\left({M_{h}\over
  1 M_\odot}\right)^{{2 \over 3}}\left({1+z \over
  1000}\right)\left({c_{s,\infty} \over 5.74 {\rm km
  s^{-1}}}\right)^{-2}.
\end{equation}
If $\chi/2 \ge 1$ the extended dark halo accretes as a point-mass with
$M=M_{h}$. Assuming $T_{gas}=T_{cmb}$, which is a good approximation
for $z>z_{dec} \sim 100$, we have approximately
\[
{\chi \over 2} \approx \left({M_{h} \over M_h^{cr}}\right)^{2 \over 3},
\]
where $M_{h}^{cr}=5196$~M$_\odot$. 

In \fig~\ref{fig:t_masses} we compare the dimensionless accretion rate
at $z=1000$ for a point mass (dashed curves) and an extended halo
(solid curves) as a function of their mass. The left panel shows the
case in which the effective gas viscosity is dominated by the Hubble
flow and the right panel shows the opposite case in which Compton drag
dominates. The results for an extended dark halo shows that the
accretion rate peaks at $M \sim M_{h}^{cr}$.  This can be easily
understood due to two competing effects. The gas effective viscosity
becomes larger with the increasing mass of the accreting object, thus
the accretion eigenvalue $\lambda$ is reduced as the mass
increases. However, the accretion rate eigenvalue for an extended dark
matter halo is increasingly reduced with respect to the point mass
case as the halo mass decreases. If $M_h \simgt 10^4$~M$_\odot$ we
have that $r_B > r_{ta}$. We will show in \S~\ref{ssec:infall} that in
this regime the Bondi solution becomes invalid and the gas flow
transitions to the time-dependent self-similar solution described by
the theory of secondary infall.

The ratio between the accretion rate onto a PBH clothed with a dark
halo and the same PBH without dark halo (``naked'') is approximately
\[
{\dot M_h \over \dot M_{bh}} = {\rm Max}\left\{\left({M_{bh} \over M_h^{cr}}\right)^{4p \over 3(1-p)}\left[{1 \over \phi_i}
     \left({1+z \over 1000}\right)\right]^{-{(6-2p) \over 3(1-p)}}, 1\right\},
\]
and is very sensitive to the slope of the halo density profile.
At a given redshift $z$, the effect of the halo growth becomes
important in increasing the accretion rate if PBHs have a mass
$M_{PBH}>M_h^{cr} [(z+1/1000)/\phi_i]^{(3-p)/2p}$. For example,
assuming a steep dark matter density profile with $p=0.2$, any PBH
``clothed'' in its dark halo, with a mass $M_{PBH}>2.37~{\rm M}_\odot
[(3/\phi_i)(1+z)/1000]^{7}$ accretes significantly more and its
accretion rate increases rapidly with time ($\propto (1+z)^{-7/3}$)
with respect to a ``naked'' PBH of the same mass. As noted in
\S~\ref{sec:dm}, density profiles steeper than $\alpha=2.25$ may be
used to describe a halo with $M_h \sim M_{pbh}$, in order to take into
account the potential of the central PBH. Thus, at redshifts $z \simlt
500$, when $M_h \gg M_{pbh}$, the fiducial value of the density
profile slope is $\alpha=2.25$ according to the secondary infall
theory. In this case ($p=0.75$ and $\phi_i=3$) the effect of the dark
matter halo becomes significant for relative massive PBHs:
$M_{pbh}>1000~{\rm M}_\odot [(1+z)/1000]^{1.5}$, or $M_{pbh} \simgt
32$~M$_\odot$ at redshift $z \sim 100$.

\subsection{Transition to time-dependent secondary infall solutions}\label{ssec:infall}

The cosmological Bondi solution rests on the assumption of steady and
non-self-gravitating spherical gas accretion. In comoving coordinates
the gas is at rest at infinity with sound speed, $c_{s, \infty}$.
Conversely, secondary infall solutions describe self-similar and
time-dependent spherical accretion in an expanding universe.  They
depend on the assumption that the matter is self gravitating and its
sound speed at infinity is $c_{s, \infty} \rightarrow 0$ (``cold''
cosmological accretion). Thus, the regime of validity of secondary
infall solutions is complementary to the Bondi solutions.  The
accretion is effectively ``cold'' if the sound speed at infinity is
much smaller than the Hubble flow: $c_{s, \infty} \ll r_b H$. This
condition is equivalent to $t_{cr}<t_H$; that is, when the hypothesis
of quasi-steady flow fails for the Bondi problem.  As shown in
\S~\ref{ssec:pm}, for an accreting point mass the transition from
steady to self-similar solutions is for $M_{pbh} \simgt 2 \times 10^4$
M$_\odot$.  This case is of little interest because the formation of
PBHs with such a large mass is unlikely. Moreover, we show later that
the accretion rate is super-Eddington, thus regulated by complex
feedbacks, the modeling of which is beyond the scope of this paper.

When PBHs do not constitute the bulk of the dark matter a dark halo
accumulates around them. Assuming, for simplicity, $T_{gas}=T_{cmb}$
we find that if $M_h \simgt 10^4$ M$_\odot$ the Bondi radius is larger
than the dark matter turnaround radius: $r_B > r_{ta}$. This result is
clearly incorrect because the gas cannot be accreted from radii larger
than the dark matter turnaround radius and is simply a consequence of
neglecting the Hubble expansion. This suggests that the Bondi solution
fails in this regime and the gas flow becomes time-dependent. Indeed,
when $r_B=GM_h/c_{s,\infty}^2 > r_{ta}$ the halo circular velocity is
larger than the gas sound speed, $(GM_h/r_{ta})^{1/2}=v_{cir}>
c_{s,\infty}$. Thus, the Bondi solutions transition to ``cold''
cosmological gas infall, for which self-similar solutions exist.  In
addition, since Hubble velocity at the turnaround radius equals the
halo circular velocity, $r_{ta} H \sim v_{cir}$, we also have $t_H <
r_{ta}/c_{s,\infty} < t_{cr}$. The assumption of steady flow is
therefore no longer correct and the flow becomes self-similar and
time-dependent.

Self-similar infall solutions for a mixture of gas and dark matter
(assuming that the gas is a small fraction of the dark matter) are
almost identical to the case of cold gas infall. The main difference
is that well behaved solutions exist for $\gamma \le 4/3$ because the
gravitational instability, leading to collapse for $\gamma \le 4/3$,
occurs only if the gas is self-gravitating.  Two types of self-similar
solutions exist. If the central black hole absorbs all the infalling
gas, the solution has free-fall boundary conditions at the center (see
discussion in \S~\ref{sec:dm}). Most of the halo gas is accreted onto
the central black hole, hence the accretion rate in units of the
Eddington rate is $\dot {m}=\dot{M}/\dot{M_{Ed}} \sim
t_{Ed}/t_H$. Since $t_{Ed} \sim 10^7$~yrs, $\dot{m} \gg 1$ at any
redshift $z \simgt 30$. The second type of solution is characterized
by shocked accretion of collisional gas. The boundary conditions are
$v=0$ at the center (\ie, no accretion onto the central black hole)
and pressureless gas upstream of the shock. A standing shock occurs at
a fixed fraction of the turnaround radius $r_s/r_{ta}$, whose value
depends on the polytropic index $\gamma$. For $\gamma=5/3$ the shock
is at $r_s=0.339r_{ta}$, analogous to the result for collisionless
matter.  For smaller values of $\gamma$ the shocked core is
smaller. Self-similar solutions can be found also for the cases in
which the central black hole accretes an arbitrary fraction of the
infalling gas. In this case the size of the shocked core is further
reduced. \cite{Tsuribe:95} found a self-similar solution for spherical
accretion around a point mass including radiation drag. They conclude
that radiation drag (i) slows down the infalling gas velocity far from
the center, (ii) flattens the density profile near the center, and
(iii) reduces the mass-accretion rate. The reduction is about a factor
of two if the gas becomes fully ionized by redshift $z \sim 400-200$.

Taking into account all the possible accretion regimes described above
may be quite complex \citep{Loeb:93, FukeU:94, Mineshige:98,
Ciotti:04,Wang:06, Park:07}. Fortunately, it is not necessary to
estimate in detail the accretion rate for the cases in which the flow
is time-dependent. This is because it is easy to show that the
accretion rate is always greater than the Eddington rate. Feedback
effects are likely to regulate the accretion rate in this
regime. Observations of quasars at moderate redshifts and theoretical
arguments \citep{Ciotti:01} suggest that when the gas supply is large
($\dot m >0.1-1$), massive black holes shine at nearly the Eddington
rate during a small fraction $\sim 1-5\%$ of their life, while during
the rest of the time they are quiescent. Most of the energy is emitted
during the active phases. Theoretical modeling of feedback processes
that produce the duty cycle is beyond the scope of the present
paper. We can treat this case phenomenologically assuming a fiducial
value of the duty cycle inferred from observations.

In order to demonstrate that when $t_{cr}>t_H$ we have $\dot{m} \gg
1$, it is convenient to rewrite the Bondi accretion formula in
\eq~(\ref{eq:dim}), in terms of the crossing time and Hubble
time. Using the relationship
$t_H^{-2}=(\Omega_{dm}/\Omega_b)G\rho_{gas}\sim 5G\rho_{gas}$, and
assuming $t_{cr} \simgt t_H$ and $\lambda \sim 1$, we have
\[
\dot m \sim \lambda {t_{Ed} t_{cr}\over 5 t_H^2} \simgt {t_{Ed} \over
  5 t_H} \gg 1,
\]
at any redshift $z>30$. Thus, as anticipated, when the accretion flow
transitions to a time-dependent self-similar regime, the accretion
rate is super-Eddington.

\subsection*{ACKNOWLEDGMENTS}
Many thanks to Jerry Ostriker for suggesting ideas that
motivated this paper and the anonymous referee for useful feedback.

\bibliographystyle{/Users/ricotti/Latex/TeX/apj}
\bibliography{/Users/ricotti/Latex/TeX/archive}

\label{lastpage}
\end{document}